# Regulatory Science Innovation for Generative AI and Large Language Models in Health and Medicine: A Global Call for Action

Jasmine Chiat Ling Ong[1,2], Yilin Ning[3], Mingxuan Liu[3], Yian Ma[3], Zhao Liang[4], Kuldev Singh[5,6], Robert T Chang[5,6], Silke Vogel[7], John CW Lim[7], Iris Siu Kwan Tan[8], Oscar Freyer[9], Stephen Gilbert[9], Danielle S Bitterman[10,11], Xiaoxuan Liu[12,13], Alastair K Denniston[12,13], Nan Liu[3,14,15+]

1.  Division of Pharmacy, Singapore General Hospital, Singapore
2.  Duke-NUS Medical School, Singapore, Singapore
3.  Centre for Quantitative Medicine, Duke-NUS Medical School, Singapore, Singapore
4.  Department of Bioengineering and Therapeutic Sciences, University of California, San Francisco
5.  Stanford University School of Medicine, Stanford, California, USA
6.  Byers Eye Institute, Stanford University, USA
7.  Centre of Regulatory Excellence, Duke-NUS Medical School, Singapore
8.  Artificial Intelligence Office, Singapore Health Services, Singapore, Singapore
9.  Else Kröner Fresenius Center for Digital Health, TUD Dresden University of Technology, Dresden, Germany
10. Artificial Intelligence in Medicine (AIM) Program, Mass General Brigham, Harvard Medical School, Boston, MA, USA
11. Department of Radiation Oncology, Brigham and Women's Hospital/Dana-Farber Cancer Institute, Boston, MA, USA
12. College of Medicine and Health, University of Birmingham, Birmingham, UK
13. University Hospitals Birmingham NHS Foundation Trust, Birmingham, UK
14. Programme in Health Services and Systems Research, Duke-NUS Medical School, Singapore, Singapore
15. NUS Artificial Intelligence Institute, National University of Singapore, Singapore, Singapore

[+]Corresponding Author

Correspondence to:

Nan Liu, Centre for Quantitative Medicine, Duke-NUS Medical School, 8 College Road, Singapore 169857, Singapore

Email: liu.nan@duke-nus.edu.sg

**Summary**

The integration of generative AI (GenAI) and large language models (LLMs) in healthcare presents both unprecedented opportunities and challenges, necessitating innovative regulatory approaches. GenAI and LLMs offer broad applications, from automating clinical workflows to personalizing diagnostics. However, the non-deterministic outputs, broad functionalities and complex integration of GenAI and LLMs challenge existing medical device regulatory frameworks, including the total product life cycle (TPLC) approach. Here we discuss the constraints of the TPLC approach to GenAI and LLM-based medical device regulation, and advocate for global collaboration in regulatory science research. This serves as the foundation for developing innovative approaches including adaptive policies and regulatory sandboxes, to test and refine governance in real-world settings. International harmonization, as seen with the International Medical Device Regulators Forum, is essential to manage implications of LLM on global health, including risks of widening health inequities driven by inherent model biases. By engaging multidisciplinary expertise, prioritizing iterative, data-driven approaches, and focusing on the needs of diverse populations, global regulatory science research enables the responsible and equitable advancement of LLM innovations in healthcare.

**Introduction**

The rapid integration of generative artificial intelligence (GenAI) in healthcare has exposed a significant regulatory gap. Generative AI algorithms are now capable of producing new text, audio, image and video outputs from training data. Examples of these algorithms used in healthcare applications include generative adversarial networks (GAN), variational autoencoders (VAE), diffusion models and transformer-based models. The transformer architecture forms the foundation for large language models (LLMs). Despite a notable increase in healthcare AI applications, existing health products regulatory frameworks across jurisdictions differ and are not specifically tailored to address the unique challenges posed by GenAI and LLMs. Current regulatory approaches were developed for AI models purposefully designed to generate specific healthcare decisions or recommendations. In contrast, LLM models are capable of a broad range of functions such as summarization of general medical information and suggesting differential diagnoses and are often applied in task for which they were not purposefully designed. This represents a shift from intended use-specific medical decision-making, which was relatively straightforward to regulate, to more generalized and flexible applications of AI.

To enable timely, safe and high accessibility to effective AI/machine learning (ML)-based medical devices, regulators including the US Food and Drug Administration (FDA) and UK Medicines and Healthcare Products Regulatory Agency (MHRA) are adaption the Total Product Life Cycle (TPLC) approach for these models.[1,2] This approach was also emphasized in the executive summary of FDA's Digital Health Advisory Committee Meeting, forming the bedrock of the committee's recommendations for regulating GenAI-enabled medical devices.[3] TPLC broadly encompasses the evaluation and monitoring of a software product from its premarket development to post-market performance, that enables rapid cycle of product improvement with safeguards.

The poor fit between LLM-based medical devices and current regulatory frameworks leaves residual risks which necessitates oversight from stakeholders across the ecosystem aside from regulatory authorities.[4] In a recent commentary, the FDA alluded to the need to keep a view on the broader ecosystem such as maintaining healthy competition among start-ups, big tech and academia; reinforcing responsibility of regulated industries; and emphasized the importance of global harmonization of regulatory standards and data definitions. In this perspective, we provide an in-depth discussion on the outstanding challenges that are not adequately addressed by a TPLC approach, focusing on regulation of LLM-based medical devices. We explore the emerging role of international and multidisciplinary regulatory science consortiums in advancing LLM innovations responsibly, as well as proposing a call for global action to drive equitable and responsible GenAI governance.

**The Constraints of the TPLC Approach**

Substantial differences exist between LLMs and AI-technologies that are already part of approved medical devices, creating unique challenges for regulation.[5] First, LLMs are trained on extensive datasets gathered from the internet and other diverse sources, making it virtually impossible to thoroughly examine or scrutinize the training data. Second, long form LLM outputs are subject to concerns over poor repeatability even with the same prompt strategy. Decision support systems using LLMs may experience performance deterioration when used in real-world settings particularly if LLMs are sensitive to changes in syntax of input queries. Third, the risk of privacy and confidentiality breaches remains unresolved. In the following section, we discuss the limits of regulatory oversight using the TPLC approach for LLMs. Figure 1 shows different phases of TPLC, and regulatory considerations specific to LLMs.

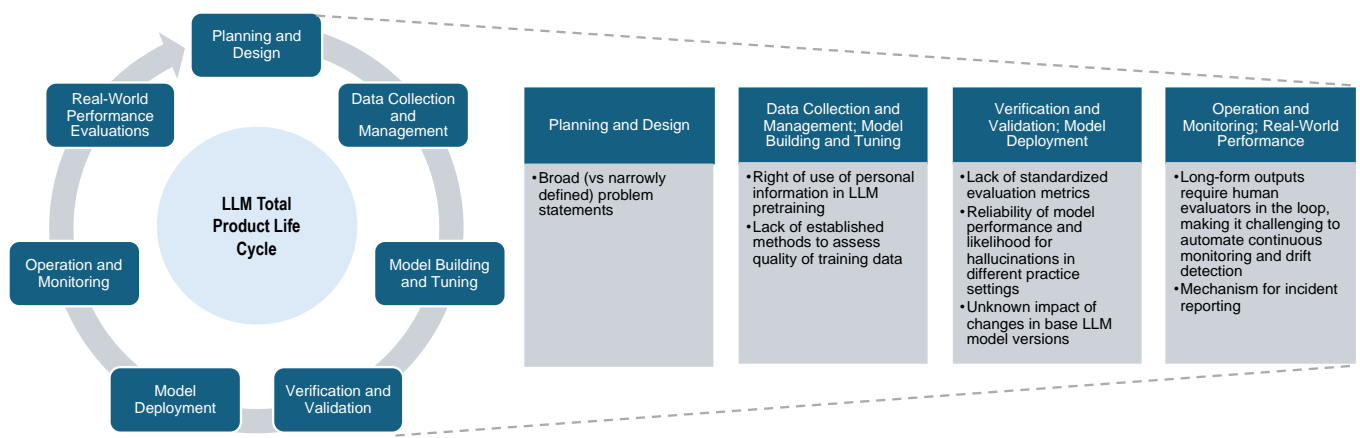

Figure 1: An illustration of the different phases of TPLC for LLM-based medical devices, and the unique considerations and regulatory challenges at each phase.

Ambiguity in Medical Device Definitions for LLMs and LLM-based applications

The qualification of LLMs and LLM-based applications as medical devices remains a topic of debate. The applicability of various regulatory frameworks, including the European Union's Medical Device Regulation (MDR) and US FDA's Software as a Medical Device (SaMD), is based on the AI model's or AI-based device's intended use and proposed indications.[6] For the EU, that means that the regulation of such LLMs under the MDR is unclear, unless they have a clear intended medical purpose.  They are considered as generative AI models under the EU AI Act but only face substantive regulation if they are classified as being of high risk or systemic risk. The downstream LLM-based applications with a medical purpose, however, would qualify as medical devices under the MDR and, if MDR risk class IIa or higher,  as high-risk AI systems under the EU AI Act. Table 1 summarises how different regulatory frameworks define a medical device. Many LLMs and LLM-based applications will likely fall into a grey area where their classification remains unclear. The description of the intended use and indications of use is less straightforward for LLMs due to their generalist nature and broad utilities. One example is LLM tools built to transcribe and summarize patient-physician interactions into a clinical document. Creation of the clinical summary requires the model to interpret clinical data and suggest medical information that may influence decisions made by subsequent reviewing clinicians. Various companies have marketed a variety of clinical documentation automation products without the need for regulatory approval as medical device. The commercial presence of these products may signal a need for stricter enforcement.

Unlike predictive models, LLM outputs are non-deterministic in nature, even when the model temperature is set to zero. Frequency of hallucinations by LLMs under different conditions of use is poorly quantified and require further examination and documentation. The risk classification and degree of control required for LLM-based medical device may need revision to cater for risks specific to LLM-based devices.

| Region | General Regulatory Pathway for Medical Devices (AI-Specific Guidance/Legislation) | Medical Device Definition | Risk Classification and Regulatory Controls | Meet Criteria as Medical Device? (Example of clinical documentation scribe) |
|---|---|---|---|---|
| USA | Food, Drug, and Cosmetic Act (FDA AI/ML-based SaMD Action Plan) | Software, when intended to treat, diagnose, cure, mitigate, or prevent disease or other conditions, are medical devices under the FD&C Act, and called "Software as a Medical Device" (SaMD) <br><br>*Software functions intended (1) for administrative support of a health care facility, (2) for maintaining or encouraging a healthy lifestyle, (3) to serve as electronic patient records, (4) for transferring, storing, converting formats, or displaying data, or (5) to provide certain, limited clinical decision support are not medical devices and are not subject to FDA regulation. | Class I (low-risk) <br> Class II (medium-risk) <br> Class III (high-risk) <br><br> Class I and II devices are often cleared through the 510(k) pathway in which applicants submit a premarket notification to demonstrate safety and effectiveness and/or substantial equivalence of their proposed device to a predicate device. <br> Class III (high-risk) devices are defined as those that support human life, prevent impairment of health, or pose potential unreasonable health risk(s) | No. |
| UK | Medicines and Medical Device Act (MHRA: Software and AI as a Medical Device Change Program, Regulation Horizons Council: The Regulation of AI as a Medical Device) | An article, instrument, apparatus or machine that is used in the prevention, diagnosis or treatment of illness or disease, or for detecting, measuring, restoring, correcting or modifying the structure or function of the body for some health purpose. <br><br>*Decision support software is usually considered a medical device when it applies automated reasoning such as a | Class I (low-risk) <br> Class II (medium-risk) <br> Class III (high-risk) <br><br> Most SaMD and AIaMD are Class II or Class III devices. Conformity assessments are required prior to receiving UKCA (UK) mark. | Probable. |

| | | simple calculation, an algorithm or a more complex series of calculations. | | |
|---|---|---|---|---|
| Europe | Medical Device Regulation (EU AI Act) | Software intended for the following medical purposes: diagnosis, prevention, monitoring, prediction, prognosis, treatment or alleviation of disease, diagnosis, monitoring, treatment, alleviation of, or compensation for, an injury or disability, investigation, replacement or modification of the anatomy or of a physiological or pathological process or state, providing information by means of in vitro examination of specimens derived from the human body, including organ, blood and tissue donations, and which does not achieve its principal intended action by pharmacological, immunological or metabolic means, in or on the human body, but which may be assisted in its function by such means. | Class I (low-risk) Class IIa (medium-risk) Class IIb (medium-risk) Class III (high-risk) AIaMD would be classified following Rule 11 of the MDR (classification rule for SaMD). AIaMD are also subject to the EU AI Act and if Class IIa or higher, are considered as high-risk AI systems under article 6. Some healthcare-related AI systems that do not qualify as medical devices fall under some provsions (e.g., transparency) of the EU AI Act, but are not generally subject to conformity assessment procedures. | Possible. |

*Table 1: Broad definitions of 'medical device' are similar across different regions. However, regulations for AI-based medical devices and software as medical devices display nuanced differences. Risk classification differs across different regions, revealing a lack of international consensus.*

Lack of Robust Evaluation for LLM Performance

We are faced with unprecedented challenges as a result of the unique capabilities of LLMs in open-ended output generation. Unlike diagnostic tests, where accuracy can be assessed, or interventional trials, where causal effects can be quantified, the evaluation of LLMs often relies on subjective and largely qualitative assessments. Multiples studies have evaluated the performance of LLMs on isolated medical tasks such as information extraction from clinical notes or answering patient enquiries. The performance of LLMs in such tasks have largely been tested in structurally simple case vignettes, medical licencing exams or medical question and answer datasets.[7] However, there is increasing recognition that the strong performance of LLMs in these tasks is not reflective of these models' performance in real-world clinical decision making. Hager et al. evaluated leading open-access LLMs using a comprehensive evaluation framework. LLMs were assessed for their performance in making diagnosis, gathering information and providing guideline adherence recommendations using real-world cases.[7] The results revealed significant limitations of LLMs, with all models performing worse than clinicians in making diagnosis, interpreting laboratory results and adherence with established treatment guidelines.

Robust evaluation of LLM bias is similarly critical to surface risks that may perpetuate health inequities. Before LLMs can be approved and used widely in clinical settings, contextualized evaluations with setting up of suitable guardrails is necessary to ensure patient safety. Various approaches to LLM bias evaluation have been proposed. One approach is to rely on physician evaluation of different dimensions of LLM bias including inaccuracy across axes of identity, lack of inclusion, stereotypical language or characterization, omission of structural explanations for inequity, failure to challenge a biased premise and potential for disproportionate withholding of opportunities or resources.[8] Another approach involves adversarial testing to probe for different failure modes and conditions whereby LLMs will generate harmful and biased outputs.[9] Convergence among regulators, manufacturers and researchers is necessary to develop evaluation methodology and metrics that are clinically relevant. The complexity of developing benchmarks depends on the deployment scenario of the LLM or LLM-based application. It could be easier for applications and models with narrow clinical use, while it would be more challenging for applications with a broad intended use.[10]

Challenges to Monitoring and Regulatory Enforcement

The training of LLMs on vast and diverse datasets present challenges in monitoring and enforcement, in particular concerning data provenance. LLM models are often trained on vast, diverse, and inconsistently documented datasets, making it difficult to trace the origins and licensing of the data used. This complicates efforts to ensure compliance with regulatory standards, such as those governing data privacy, consent, and intellectual property rights. A recent large-scale audit of over 1,800 text datasets on Hugging Face revealed frequent misclassifications. License omission was reported in 70% of datasets and errors were detected in 50% of listed licences. This highlights a problem in misattribution of datasets and informed use in AI and LLM development.[11] In addition, the open-sourced models allow modifications and retraining by third parties to be performed freely, further obscuring data provenance and undermining data authenticity. Addressing these challenges necessitates the development of robust frameworks for data auditing and provenance tracking to enable more effective oversight of LLM-based tools.

Post-marketing surveillance of approved LLM-based tools has similar challenges to monitoring the effects of medicinal products post licensure[12,13]. First, the inability to track "off-labelled" use of domain specific-LLMs for example when applying the models for unapproved indications and/or patient populations[14]. Signal detection of public health risks is very challenging given the pervasiveness of LLM tools with diverse applications and uses. Second, reliance on voluntary reporting of adverse events by clinicians to potential under-reporting and delays in signal detection.[15-17] LLM-based tools, such as Ambient AI medical scribes, heighten the challenge of identifying potential adverse events attributed to use of these tools. Unedited "hallucinations" or inappropriate redaction of pertinent information may be carried forward to subsequent clinical notes, resulting in errors of unknown provenance. Finally, high cost and extensive resources are required to develop a robust vigilance program due to the diverse nature of LLM applications and unknown profile of risk to patients.

Accelerated market approval processes pose additional challenges for monitoring programs. For example, the FDA has approved close to two-third of all AI-based SaMD devices via the 510(k) pathway.[18] This process for medical devices is based on "substantial equivalence" to devices cleared pre-1976 or legally marketed thereafter, known as predicate devices.[19] A 'predicate creep' has been described as a cycle of technology change through repeated clearance of devices based on predicates with slightly different technological characteristics. While the 510(k) pathway has facilitated faster clearance of devices, these may have referenced multiple iterations of predicates, resulting in devices with significant differences in features and deviating from the design intent of the first rigorously reviewed and approved reference model; this could pose risks to patient safety.[20,21] Depending on risk assessment, some GenAI tools may be approved via the 510(k) pathway on the premise of 'low-risk, design iterations' built upon the same pre-trained foundation models such as fine-tuning or employing a retrieval-augmented generation framework. These 'design iterations' enhance performance but may also significantly change the nature of the model and therefore its risks.[22,23] From the perspective of the user, even variations in user prompts may significantly impact the reproducibility and reliability of responses from LLMs.[24,25] Some models have been cleared with predicates that were not ML/DL (deep learning) devices. It is hence possible that LLMs could be cleared with predicates that were not LLM or even ML/DL-based.

The rising pervasiveness of the technology makes enforcement challenging. Growing evidence suggests that open-sourced LLMs are fast approaching performance of proprietary LLMs in medical tasks[26,27]. As a result of dramatic cost reductions per token of proprietary LLMs such as GPT-4o by OpenAI[28], the access barrier is lowered and we anticipate widespread adoption of LLMs for both medical and non-medical purposes in the absence of formal approval and/or robust evidence, posing challenges for enforcement. Developers continue to release unapproved, public-fronting LLM-based applications, despite meeting EU and US criteria of a medical device.[10,29]

Ethical Considerations Beyond Current Regulatory Oversight

The widespread availability and rapid adoption of LLM tools, either directly applied to or adapted for healthcare, are raising numerous ethical dilemmas and posing potential risks to the broader public. Risks that remain inadequately addressed include low trust of LLM-based health applications due to risks of hallucinations and poor reproducibility of output; embedded bias in LLMs exacerbating health inequities; and bioethical concerns such as patient privacy arising from the use of healthcare data in model pre-training.[30-34] Despite its promise, current evidence is still lacking on the impact of LLMs on patient outcomes. In addition, the impact of LLM applications on population health and minority populations is still largely unknown.

Regulatory guidance for responsible and ethical LLM use is critical. The issue of bias being demonstrated or exacerbated by AI algorithms is an increasing concern within the healthcare industry. Several initiatives such as the STANDING Together initiative (standards for data diversity, inclusivity, and generalizability)[35] and the FDA's "Artificial Intelligence/Machine Learning (AI/ML)-Based Software as a Medical Device Action Plan" serve to bridge the gaps inherent in existing research and regulatory standards/guidance.[36] However, there has been little formal consideration of the impact of patient interactions with AI programs, including the impact of LLM medical tools on patient autonomy, dignity, the patient-clinician relationship and trust.[37] Conversational applications (chatbots) powered by LLMs allow dynamic, context-aware exchanges in patient education. However, LLM interactions are nuanced, thus model fine-tuning is typically needed, and requires post launch monitoring and policy optimization to reduce risks of embedded cognitive biases such as confirmation bias, automation bias and automation complacency as well as avoiding model drift.

Developing effective regulatory policies which take into consideration the vast number of ethical principles is challenging. Interpretation of ethical principles can be highly subjective, and may vary in different contexts of LLM application. In addition, most ethical guidance offers little assistance when trade-offs need to be made between ethical principles. To address these challenges, a multidisciplinary approach involving bioethicists, regulators, users and manufacturers is required.

**Regulatory Science in LLM Innovation – Opportunities and Trends**

Regulatory science is the science of developing new tools, standards, and approaches to assess the safety, efficacy, quality, and performance of health products that are assessed by regulatory agencies.[38] Regulatory authorities such as the FDA, European Medicines Agency (EMA), Australian Therapeutic Goods Administration (TGA) and the Singapore Health Sciences Authority (HSA) conduct research or partner with regulatory research groups in the development of guidance material. Such regulatory collaborative groups include Centres of Excellence in Regulatory Science and Innovation in the United States, Centre for Regulatory Science and Innovation in the United Kingdom, the Duke-NUS Centre of Regulatory Excellence in Singapore; and at a global level, the Global Coalition for Regulatory Science Research and World Health Organization (WHO). In this section, we discuss the priorities of regulatory science research in advancing the goals of responsible LLM innovation and balanced regulation.

Applying Adaptive Regulatory Approaches

Adaptive regulatory approaches may be adopted for policies characterized by a fast pace of innovation. LLM-based tools are prime candidates for this regulatory approach and could catalyse advances in this area, commanding the attention of regulators and industry stakeholders to keep up to date with growing scientific evidence of risks and benefits of the technology.[39] Adaptive approaches are those that empower regulators to be less restrictive in the absence of negative outcomes demonstrated in clinical trials, and reinstate restrictions should evidence of harm emerge in the process of post-marketing surveillance.[39] For example, accelerated approval (in the US) and conditional marketing authorization/approval (in the EU, Singapore and Japan) pathways are in place for emerging therapeutics.[40] The predetermined change control plan proposal put together by FDA, Health Canada and MHRA supplements the TPLC, allowing certain changes that were predicted and predefined in a submitted plan, to be made without requiring another approval process; these changes could be a result of information from real-world use. A more flexible approach is an anticipatory regulatory approach, whereby regulatory rules are iteratively developed alongside development of new products or services. Regulatory sandboxes are outcomes-oriented tools used to guide anticipatory regulation, whereby new services, health products or digital health tools can be tested in a constrained environment with less regulatory requirements. Examples include the United Kingdom's MHRA AI Airlock and Singapore's Infocomm Media Development Authority (IMDA) Privacy Enhancing Technology Regulatory Sandbox.[41,42] Similarly, the EU AI Act calls for establishing regulatory sandboxes for improving regulatory compliance with the act (detailed in Article 57 of the EU AI Act). These sandboxes allow developers and researchers to work closely with regulators to identify, quantify and prospectively develop solutions to mitigate potential risks in early stages.

Regulatory authorities can work with local and global regulatory research groups in the conceptualization, conduct and assessment of outcomes from such regulatory sandboxes. Such a setup enables regulatory experimentation to gain rapid understanding on costs and benefits of different frameworks and policies. The main characteristics of regulatory sandboxes include the following: (1) they are adaptive; (2) they require close collaboration and iteration between stakeholders such as users, regulators and industry; and (3) they are evidence-based, with a trial and error approach.[43] "Global regulatory sandboxes" can be developed to study the international interoperability of regulatory policies and the impact of policies on cross-border innovation and competition.

Convergence of Regulatory Frameworks: Medical Device and Health Services

AI agents powered by LLMs and agentic systems show promise in enabling greater level of automation in clinical tasks.[44] AI agents in healthcare are intelligent systems designed to assist, automate, and enhance various aspects of medical care. Agentic systems have demonstrated improved capabilities over single LLM models in medical information processing, planning and making clinical or operational decisions through interacting and collaborating among different agents.[45] These capabilities are shifting LLM agents from being pure assistive tools into service agents with elevated levels of autonomy, much like a skilled worker performing a service. The development of AI agents powered by LLMs, in particular agents with reasoning capabilities,[46] is introducing a paradigm shift. For example, OpenAI o1 model was designed to answer complex questions through a series of intermediate reasoning steps, a strategy known as "chain-of-thought".[47] The scaling of automation beyond repetitive, low skill tasks lends in to a 'Software as a Medical Service' (SaMS) model. Coupled with high accessibility to cloud-computing resources, implementation barriers and costs to health systems of AI-based software and tools is significantly lowered. This allows healthcare providers and patients to access cutting-edge tools including AI agents without the need for extensive upfront infrastructure investments.

As opposed to a tangible product with defined functionalities, SaMS is focused on service-oriented delivery, much like a digital health service. In Singapore, health services regulation is coming under the ambit of HSA which had previously only regulated health products; this will result in greater convergence of products and services regulation in future, providing a more optimal framework for regulating healthcare AI.[48] Such novel regulatory frameworks will need to be co-developed with regulatory science groups and studied before LLM agents and systems can be adopted at scale.

Increase Focus on AI Supply Chain

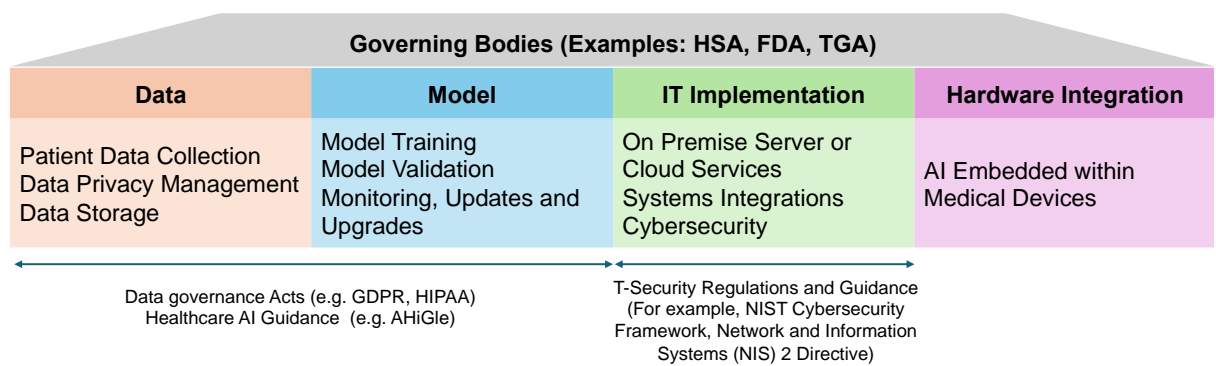

*Figure 2: An illustration of the Healthcare AI Supply Chain with examples of governing bodies and regulations.*

The recent global IT outage incident that started with a single software update by the cybersecurity company CrowdStrike is a prominent example of how a single failure point in the digital delivery system can lead to a global disruption of critical health services including individual products.[49] Widespread digitization, adoption of electronic medical records, mHealth and proliferation of AI-based tools add to our dependency on digital solutions and growing complexity which cannot afford any disruptive down time. Beyond predictive, diagnostic and decision support algorithms, it is important to recognize that AI is likely to be pervasive in operating systems, software and even embedded within hardware such as computer chips.[50] Ensuring continuity of care delivered to patients requires a scrutiny of vulnerabilities in the use of AI across the process of provision of healthcare delivery, both in the business and service value chain. Figure 2 shows the healthcare AI supply chain and examples of governance and guidance frameworks.

A global collaborative effort in regulatory science enhances and strengthens the following: internal and global visibility on various components of the supply chain; frameworks to guide identification, testing and validation of critical components and pipelines; sharing good practice for resilience teams and business continuity plans (BCP); facilitate communications strategy across supply chain networks. We also propose this global effort to support health systems build AI development capabilities, develop infrastructure for local deployment (on premise or edge) of these models. This serves as a fail-safe strategy, reducing dependency on commercially available, leading technologies. In addition, such deployment strategies improve availability of services and lower dependence from commercial cloud-based products. This approach could lower privacy risks while having its own cybersecurity challenges. There is a need to ensure that capacity building and participation in a global collaborative effort is not limited to researchers from high income countries and representing top-tier research institutions.[30] Along the same line, continued efforts to improve open-sourced models (e.g. open-sourced LLMs) aligns with commercial competition laws.[51] Government policies and health system investments need to align with these strategies, emphasizing diversity in technology investments to build a more resilient healthcare AI infrastructure.

**Future Directions for Regulatory Science and Regulators**

Beyond Medical Device Regulation: Responsible AI in Health Product Development

AI has significant potential to influence various aspects of medical product development, and this impact is already in progress. The drug development industry has undergone a revolution with introduction of computational approaches including deep learning, foundation models and quantum computing.[52] Deep generative AI has shown huge promise in de novo molecular design and early development stages e.g. predict protein folding, molecular interactions, and cellular disease processes.[53,54] ADMET (Absorption, Distribution, Metabolism, Elimination and Toxicity) properties of small molecules can now be predicted with high degree of accuracy.[55] As these AI-designed targets move into clinical testing, the industry is embracing another revolution in the clinical research process to accelerate and streamline the process of clinical trials. In particular, the advent of LLMs has brought about rapid advancements in AI-aided clinical trial design, participation selection and recruitment as well as tracking and promoting adherence to trial protocols.[4] Such applications present immense promise in shortening the cost and development time for new drug entities. However, unknown risks associated with using the technology to augment or automate critical processes warrants close scrutiny by the research community and regulators.

We believe that LLMs can bring about significant value in reducing administrative burden within regulatory agencies, and increase sensitivity and reliability of post-marketing surveillance and evaluation. Studies suggest promising performance in early identification and classification of adverse drug events and automation of drug approval processes.[56] Such tools are of low clinical risk, but bring about significant returns in reducing documentation burden and turnaround time for new drug applications. More funding and collective support for research in this area will bring about benefits in accelerating medical product time to market.

Global Call to Action

The International Medical Device Regulators Forum (IMDRF) is an example of a global effort to promote harmonization medical device regulations including ML/AI-enabled devices.[57] The committee is represented by regulatory authority representatives from different jurisdictions including Australia's TGA, Health Canada, Singapore's HSA, the EU's European Commission - Directorate-General for Health and Food Safety and the US FDA. Global regulatory research organizations, such as the Global Coalition for Regulatory Science Research and the Global Network for Regulatory Science, are pivotal in advancing regulatory research through fostering international collaboration. By engaging with industry and regulatory stakeholders from various regions, they play a crucial role in establishing cross-industry, cross-national conversations, global harmonization and drive outcomes research to facilitate formulation of regulatory guidelines for responsible adoption of AI and LLMs. The worldwide, interconnected nature of digital health and AI technologies underscores the importance of adopting a global perspective and collaboratively addressing pressing issues at hand.

One such pressing issue is the lack of robust standardization and harmonization for LLM based tools. A recent literature review of 142 LLM studies in medical applications revealed significant gaps in overall reliability, generalizability, and applicability of model evaluation practices.[58] In a separate systematic review concerning current AI reporting and evaluation methods, 220 distinct items were recommended for detailing model performance or describing source data characteristics through "model cards" and "data cards".[59] "Model cards" encourage transparent reporting of AI model performance characteristics while "data cards" describe the diversity of AI model training datasets.[60] Several groups have attempted to address this gap, notably CONSORT-AI, SPIRIT-AI, TRIPOD-AI, PROBAST-AI, MI-CLAIM, and most recently, TRIPOD-LLM for transparent reporting of LLMs in scientific publications.[61-63] The TREGAI and CARE-AI checklists addresses ethical reporting of AI models[64,65] while health AI assurance labs such as CHAI, VALID-AI and HAIP[66,67] facilitates private-public partnerships and booster collaborative governance of medical AI models at a local level.[66,68] As these different checklists and governance groups start to form, alignment in directions and consolidation of functions will be critical. Global regulatory research groups can take on such a function, establishing common grounds for conversations among different stakeholders while allowing room for localization of regulations.

Advancing the Collective Goals of Heath Equity

Global divides between high-income countries (HICs) and low- and middle-income countries (LMICs) are leading to disturbing health inequities.[69,70] Regulatory bodies and international organizations are taking actions to push health equity to the forefront in regulatory strategies, e.g. the World Health Organization global Guidance for best practices for clinical trials to promote equitable clinical trials, and the FDA's Health Care at Home Initiative to drive equity in digital medicine.[71] LLMs have the potential to both bridge and exacerbate health inequities. Potential harms can arise if social and structural health determinants; geographic, linguistic, and demographic biases in datasets[72], misconceptions tied to patient identity; prioritization of privileged perspectives; and systemic disparities in system performance and accessibility across populations are not appropriately addressed.[8] There is also a need to identify and understand the various nuanced failure modes of LLM tools when implemented at scale, to allow health delivery systems to devise more effective real-time monitoring programs. For example, repeated testing (conducted over a thousand times) of LLMs using hypothetical clinical scenarios revealed clear gender and racial biases.[73] These may not be apparent when tested in a clinical trial, but emerge as critical public health concerns when deployed in a real world setting. LLMs and GenAI risk perpetuating biases in clinical applications at production scale with biases accumulating throughout data processing and modelling process.

Global regulatory science groups can advance the goals of health equity beyond country borders and geopolitical regions. Intentional inclusion of LMIC perspectives in global discussions to tailor AI solutions that address specific regional health challenges and resource constraints. The goal of such efforts is to democratize resources and extend support to regions and ecosystems with limited access. Knowledge and resource transfer enables accessible and affordable digital and AI interventions that are safe, and effective. Translational gaps exist for innovative technologies in LMICs, primarily arising from complex economic, infrastructural, and political barriers.[74] Economic and political interests from industry, nongovernment organizations and private funders within LMICs may drive agendas that influence regulatory decisions. AI-supported screening programs for chronic diseases such as diabetic retinopathy (DR) have reported varying degrees of success in low resource settings.[75] Various implementation and adoption barriers were reported such as the inability to provide timely reports and lack of patient adherence to program recommendations.[76,77] Research on LLMs in LMIC has demonstrated potential benefits. In one prospective study, patients randomized to the intervention arm received diabetes self-management recommendations from primary care physicians assisted by an LLM integrated tool.[78] Patients in the intervention group showed better self-management behaviour compared to patients in the control group where primary care physicians were unassisted. However, the authors note raised the need to address various implementation challenges, including data quality and barriers to integration with existing healthcare information technology infrastructure.

**Conclusion**

As GenAI and LLM continue to transform healthcare and diversify its applications, we will see greater challenges and uncertainties in development of regulatory policies. There will be a paradigm shift in how we view the TPLC approach, and a need for innovative and responsive regulatory framework. We recognize the importance of ethical considerations, services regulation and other challenges that that fall beyond the traditional boundaries and oversight of health product regulators. Adherence to ethical guidelines, maintenance of good supply chain practice policies, and effective education for both manufacturers and users will be critical for LLMs to deliver on the promise of improving global health. Independent global regulatory science research groups can play an important role in integrating insights from different sources of information and industries to ensure safe, beneficial, and equitable use of LLMs in healthcare.

34. Danielle S, B., Hugo JWL, A. & Raymond H, M. Approaching autonomy in medical artificial intelligence. *The Lancet. Digital health* **2**(2020).
35. Shaswath, G*., et al.* Tackling bias in AI health datasets through the STANDING Together initiative. *Nature medicine* **28**(2022).
36. Mirja, M., Marium M, R. & Joseph C, K. Bias in AI-based models for medical applications: challenges and mitigation strategies. *NPJ digital medicine* **6**(2023).
37. Amar H, K*., et al.* Digital Health to Patient-Facing Artificial Intelligence: Ethical Implications and Threats to Dignity for Patients With Cancer. *JCO oncology practice* **20**(2024).
38. Food and Drug Administration. U.S.A. Focus Areas of Regulatory Science - Introduction | FDA. Accessed through: https://www.fda.gov/science-research/focus-areas-regulatory-science-report/focus-areas-regulatory-science-introduction (2024).
39. Grandis, G.D., Brass, I. & Farid, S.S. Is regulatory innovation fit for purpose? A case study of adaptive regulation for advanced biotherapeutics. *Regulation & Governance* (2023).
40. H-G, E*., et al.* From adaptive licensing to adaptive pathways: delivering a flexible life-span approach to bring new drugs to patients. *Clinical pharmacology and therapeutics* **97**(2015).
41. Emily, L., Dalia, D., Jacoline, B. & Pall, J. The Sandbox Approach and its Potential for Use in Health Technology Assessment: A Literature Review. *Applied health economics and health policy* **19**(2021).
42. Infocomm Media Development Authority, Singapore. Privacy Enhancing Technology Sandboxes - Infocomm Media Development Authority. Accessed through: https://www.imda.gov.sg/how-we-can-help/data-innovation/privacy-enhancing-technology-sandboxes (2024).
43. OECD. Regulatory sandboxes in artificial intelligence. Vol. No. 356 (OECD Digital Economy Papers, 2023).
44. Qiu, J*., et al.* LLM-based agentic systems in medicine and healthcare. *Nature Machine Intelligence* **6**, 1418-1420 (2024).
45. Mehandru, N*., et al.* Evaluating large language models as agents in the clinic. *npj Digital Medicine* **7**, 1-3 (2024).
46. Introducing OpenAI o1. (@OpenAI, 2024).
47. OpenAI. Learning to Reason with LLMs. Vol. 16 December 2024 (https://openai.com/index/learning-to-reason-with-llms/, September 12, 2024).
48. John CW Lim, Tan Koi Wei Chuen & Vogel, S. Regulating the Future of Health: CoRE's Tenth Anniversary Perspective. in *CoRE Regulatory Perspective*, Vol. 2024 (Duke-NUS Medical School, Centre of Regulatory Excellence, https://www.duke-nus.edu.sg/core/think-tank/core-regulatory-perspective/regulating-the-future-of-health-core-s-tenth-anniversary-perspective, 2024).
49. Association, A.H. CrowdStrike Technology Outage Causing Global Disruption, Including Impacts on Hospitals | AHA. (@ahahospitals, https://www.aha.org/advisory/2024-07-19-crowdstrike-technology-outage-causing-global-disruption-across-industries-including-impacts-hospitals-and, 19 July, 2024).
50. Biglari A & Wei, T. A Review of Embedded Machine Learning Based on Hardware, Application, and Sensing Scheme. *Sensors (Basel)* **4**, 2131 (2023).
51. Riedemann, L., Labonne, M. & Gilbert, S. The path forward for large language models in medicine is open. npj Digit. Med. 7, 339 (2024).
52. Chiranjib, C., Manojit, B. & Sang-Soo, L. Artificial intelligence enabled ChatGPT and large language models in drug target discovery, drug discovery, and development. *Molecular therapy. Nucleic acids* **33**(2023).
53. Druedahl, L.C*., et al.* Use of Artificial Intelligence in Drug Development. *JAMA Network Open* **7**(2024).
54. Amit, G. & Antonio, L. Unleashing the power of generative AI in drug discovery. *Drug discovery today* **29**(2024).
55. Swanson, K*., et al.* ADMET-AI: a machine learning ADMET platform for evaluation of large-scale chemical libraries. *Bioinformatics* **40**(2024).
56. Ong, J.C.L*., et al.* Generative AI and Large Language Models in Reducing Medication Related Harm and Adverse Drug Events – A Scoping Review. medRxiv. Accessed: https://doi.org/10.1101/2024.09.13.24313606 (2024).
57. International Medical Device Regulators Forum. Working Groups: Artificial Intelligence/Machine Learning-enabled. Vol. 2024 (https://www.imdrf.org/working-groups/artificial-intelligencemachine-learning-enabled).
58. Thomas YC, T*., et al.* A framework for human evaluation of large language models in healthcare derived from literature review. *NPJ digital medicine* **7**(2024).